\documentclass[12pt]{article}
\usepackage{epsfig}

\usepackage{amsmath}

\def\beq{\begin{equation}}
\def\eeq#1{\label{#1}\end{equation}}
\def\eeqn{\end{equation}}

\def\beqa{\begin{eqnarray}}
\def\eeqa#1{\label{#1}\end{eqnarray}}
\def\eeqan{\end{eqnarray}}

\let\bar=\overbar

\def\Dslash{\not{\hbox{\kern-4pt $D$}}}
\def\dslash{\not{\hbox{\kern-2pt $\del$}}}

\def\msb{{\bar{\ssstyle M \kern -1pt S}}}

\def\Title#1{\begin{center} {\Large {\bf #1} } \end{center}}

\begin{document}
	
	\begin{flushright}
	UTTG-03-13
	\end{flushright}
\bigskip

\Title{New Physics and Single Top Production\footnote{Proceedings of CKM 2012, the 7th International Workshop on the CKM Unitarity Triangle, University of Cincinnati, USA, 28 September -- 2 October 2012}}

\bigskip\bigskip


\begin{raggedright}

{\it Jiang-Hao Yu\\
Theory Group, Department of Physics \\
The University of Texas at Austin \\
Austin, TX 78712 USA}
\bigskip\bigskip
\end{raggedright}

\bigskip\bigskip

\begin{center}
{\small\bf Abstract}
\end{center}
{\small
We show how single top production in conjunction with a variety of additional final states provides a sensitive probe of new physics. 
We investigate several new physics scenarios such as lepto-phobic $W'$ model,  and exotic $B'$ model, and then perform detailed simulation analyses in $s$-, and $tW$-associated production channels.
}

%
%

\section{Introduction}

The top quark plays a special role in the Standard Model (SM) because of its heavy mass, and large Yukawa coupling.
At the hadron colliders such as Tevatron and Large Hadron Collider (LHC), the top quark is
mainly produced through QCD process, while the electroweak contributions are diluted.
However, processes involving single top production are unique in their ability to probe the nature of electroweak symmetry breaking. 
 The fact that the top quark mass is of order the electroweak scale suggests that top quark production may be sensitive to new physics beyond the SM.

Although the top quark has been observed in the top pair production at the Tevatron in 1995~\cite{Abe:1995hr}, observation of SM single top production is experimentally challenging. SM single top production dominantly occurs  through electroweak processes while SM top quark pair production mainly through the strong interactions.  Therefore the single top SM production rates are smaller than the top pair production rates.

The single top production has been observed in three channels: $s$-channel, $t$-channel~\cite{Abazov:2009ii}, and associated production channel~\cite{:2012dj}, which agree with the SM predictions within experimental uncertainties.
So the new physics (NP)~\cite{Tait:2000sh, Cao:2007ea} can contribute to (a) excess on cross sections through new particles or modified couplings; 
(b) modified distributions, such as resonance effects, enhanced shapes, etc. 
We show how single top production in conjunction with a variety of additional final states can provide a sensitive probe of such NP. Using this final states, we investigate several new physics scenarios such as $W'$, $T'$, or $B'$, and then perform a detailed simulation analysis in $s$-,  and associated production channels.

\section{New Physics in $s$-channel}

In the $s$-channel process, new resonances can be produced and then decay to the $tb$ final states, which can be fully reconstructed.
Searches for $tb$ resonances have been investigated at the Tevatron and LHC~\cite{Abazov:2011xs}.   
Since the SM contribution in the $s$-channel process is smaller than the one in the $t$-channel process, 
it is more sensitive to new resonances in the $s$-channel process.
The possible resonances are charged vector boson $W'$, charged scalar $H^\pm$, exotic scalars such as color sextet, or triplet, etc.

We focus on searches for a lepto-phobic $W'$~\cite{Cao:2012ng}. 
Usually, the leptonic final state will be the primary search channel for $W'$. 
However, because $W'$ searches in the leptonic final states at the LHC are pushing to higher and higher $W'$ mass regions, 
it is also worth to search for $W'$ in the quark channel, especially single top quark channel in the low mass region.
For example, in some special models in which the $W'$ is lepto-phobic,
the third generation quark final state will be the best channel to look.
In the top-flavor model, the $W'$ prefer to couple with the third generation quarks.
To keep this analysis model independent, the most general effective Lagrangian can be written as
\begin{eqnarray}
{\cal L} =  \dfrac{g_2}{\sqrt{2}}~\left(\begin{array}{ccc}\bar{u} & \bar{c} & \bar{t} \end{array}\right)
\gamma^\mu (f_{L} P_L +  f_{R} P_R) 
\left(\begin{array}{c} d \\ s \\ b \end{array}\right)
~W^{\prime +}_\mu + {\rm h.c.}~~, 
\end{eqnarray}
where $g_2=e/\sin\theta_W$, and $P_{L/R}$ are the chirality projection operators.

Considering the semi-leptonic decay of the top quark, the final state of this channel is
\begin{eqnarray}
		p p \to W' \to t \bar{b} (\bar{t} b) \to b l \nu \bar{b}.
\end{eqnarray}
The dominant backgrounds to the final state of lepton and two jets are from top quark pair production and
$W$~boson production in association with jets. 
Smaller backgrounds are from single top quark production in association with a 
$W$~boson ($t+W$) or with jets ($t+jets$, $t$-channel and $s$-channel) and from 
diboson+jet ($WV$) production.

We use the anti-kt algorithm to cluster quarks
and gluons into final state jets. 
Detector resolution effects are simulated by smearing jet and leptonic energies.
We model $b$-tagging as a flat 60\% probability to tag $b$-quark jets and a 0.5\% probability
to mistag non-$b$-quark jets (including charm quarks).
At the analysis level, all the signal and background
events are required to pass the {\it basic} selection cuts.
To isolate the $W'$ signal and suppress the SM backgrounds, a set of final cuts is applied on
the jet $P_T$ and on $H_T$, 
where $H_{T}$ is the scalar sum of the system  $p_{T}$.
These cuts effectively suppress most of the SM backgrounds while passing much of the 
$W'$ signal. 
In order to further improve the sensitivity of the analysis, the reconstruction of the 
$W'$ and its invariant mass is required. 
For this reconstruction it is necessary to 
first obtain the neutrino momentum.
If there are two solutions of the neutrino $p_z$ momentum, we will pick up the solution with central rapidity. If the solution is complex, we will pick up the real part of this solution.
With the neutrino identified properly, we reconstruct the mass of the $W'$ as
\begin{equation}
m_{W'}^{\rm rec} = m(\vec{p}_{\nu} + \vec{p}_l + \vec{p}_{\rm jet\, 1}
               + \vec{p}_{\rm jet\, 2} )
\end{equation}
To increase the local significance of this process, after all the optimized cuts one will select certain bins of the invariant mass distributions.
We then impose a window cut on the invariant mass difference between the reconstructed
invariant mass and the theoretical  $W'$ mass.

After the discovery of the $W^\prime$, one would like to know its mass, spin, and couplings.   
The chirality of the $W^\prime$ coupling to SM fermions is best measured from the polarization of the top quark. The charged lepton from top-quark decay is maximally correlated with top-quark spin.  
Clearly, the charged lepton from a right-handed top-quark prefers to move along the top-quark moving direction 
while the one from a left-handed top-quark is to 
against the top-quark moving direction.
The angular correlation of the lepton is $\dfrac{1\pm \cos\theta_{\rm hel}}{2}$ with the ($+$) choice for right-handed and ($-$) for left-handed top-quarks, 
where $\theta_{\rm hel}$ is  the angle of the lepton in the rest frame of top quark relative to the top-quark
direction of motion in the overall c.m. frame.

\section{New Physics in $tW$-channel}

In the top quark associated production channel, the SM production rate is tiny at the Tevatron, but significant at the LHC.
Through flavor changing neutral current, NP can contribute to the following final states: $tW$, $tZ$, $t\gamma$, $tH$, and monotop, etc.  

We consider a $B'$ model~\cite{Nutter:2012an}, which preferentially 
decays into a single top quark produced in association with a $W$~boson. An effective scenario where a new $B'$~quark is the only 
light state below a cutoff $\Lambda$ is considered. 
The most general Lagrangian describing the interactions of heavy bottom quarks with 
gluons (assuming operators of dimension five or less) is~\cite{Baur:1987ga}
\begin{equation}
	{\mathcal L} =  g_s\,\overline{B'} \gamma^\mu G_\mu B' + \frac{g_s\,\lambda}{2\,\Lambda} G_{\mu\nu}\, \overline{b}\, \sigma^{\mu\nu} \biggl(\kappa^b_L P_L + \kappa^b_R P_R\biggr) B' + \mathrm{h.c.} \label{eq:productionL}
	\end{equation}
where 	$\lambda$ is a free parameter whose value is dependent on the 
	UV physics that was integrated out. 
The dimension five operator is generated in many models 
by integrating out new states.
Because of the 
large fraction of  gluon initial state partons at LHC energies, this process could have visible production rate.  Similar operators
can generate flavor-changing-neutral-currents (FCNC).  We assume the UV theory is 
free of FCNCs, therefore ensuring that $\lambda/\Lambda$ is sufficiently suppressed.

The electroweak decay of the $B'$~quark into a single top quark is parametrized as
\begin{equation}
	{\mathcal L} = \frac{g_2}{\sqrt{2}} \,W^+_{\mu} \,\overline{t} \gamma^{\mu}\, \bigl(f_L P_L + f_R P_R \bigr) \,B' + \mathrm{h.c.}
\end{equation}
Regarding to $B'$ decay branching ratio,
at low masses, 
the $bZ$ and $bH$ decays dominate, while at higher masses
the $Wt$ decay is the largest. 
The large decay
branching ratio to $Wt$ makes this an attractive final state for a $B'$~search.

We consider
the lepton+jets $B'$ final state and evaluate the backgrounds to this signature. We look at 
the leptonic decay process $p p \to B \to t \bar{b} \to b l^+ \nu \bar{b}$. 
The dominant
backgrounds to the final state of lepton and three jets are from top quark pair production and
$W$~boson production in association with jets. 
For top pair production we include both the 
lepton+jets final state, $t\bar{t} \to  b l \nu\, \bar{b} jj$, and the dilepton+jet final 
state, $t\bar{t}j \to  b l \nu\, \bar{b} l \nu\, j$. For the lepton+jets final state, one of
the jets must be at low $P_T$ or otherwise be lost in order to enter the signal region. 
For the dilepton+jet final state, one of the leptons must be at low $P_T$ or otherwise
be lost. 
Smaller backgrounds are from single top quark production in association with a 
$W$~boson ($t+W$) or with jets ($t+jets$, $t$-channel and $s$-channel) and from 
diboson+jet ($WV$) production.

Of course, signal and background events are required to pass the basic selection cuts.
The backgrounds are mostly
at low $P_T$, whereas the $B'$ signal is at high $P_T$. The top quark pair background 
extends farthest into the $B'$ signal region. This can also be clearly seen in the distribution
of $H_T$.
To isolate the $B'$ signal and suppress the SM backgrounds, a set of final cuts is applied on
the jet $P_T$ and on $H_T$.
To suppress the background from $W$+jets and dibosons further, we require at least one 
jet to be $b$-tagged. 
These cuts effectively suppress most of the SM backgrounds while passing much of the 
$B'$ signal. 
The largest remaining background contribution is from top pair production. At low $H_T$, 
$W$+jets also contributes, less so at high $H_T$. 
In order to further improve the sensitivity of the analysis, the reconstruction of the 
$B'$~quark and its invariant mass is required. 
After reconstructing the mass of the $B'$~quark, we then impose a window cut on the invariant mass difference between the reconstructed
invariant mass and the theoretical $B'$ mass. 
In particular the $b$-tagging cut reduces the $W$+jets background 
significantly.

If a $B'$ is found, then it is possible to determine if it has left-handed or right-handed
couplings by looking at the $W$~boson helicity from the top quark decay. 
At the parton level before any selection cuts, this results in
the familiar SM-like distribution for left-handed $B'$. The right-handed $B'$ distribution
is quite different, and the clear distinction remains even after selection cuts.

\section{Summary}

We investigated the sensitivity of the $s$-channel $W'$ mode,  and the $tW$-channel $B'$ mode to various possible forms of NP.
Furthermore, NP can contribute to the $t$-channel process through flavor changing neutral vector bosons $Z'$ or $G'$,
although there are strong constraints from flavor physics. 
In the $t$-channel, it can also be resonance production through $p p \to T  q$ or $p p \to B  q$, where
$T$ or $B$ are heavy quarks.
The three modes are sensitive to different forms of NP, which provide complimentary information about the top quark properties. 
We mainly focus on resonance productions in various channels, which could significantly contribute to excess on cross sections and modify invariant mass distributions.
Although current data show there is no hints on NP, 
it is still useful to set the upper limits on masses of such heavy resonances, or the effective couplings in the effective theory approach.


\bigskip
{\it Acknowledgement:} I am grateful to C.-P. Yuan, Devin G.E. Walker, Reinhard Schwienhorst and Joseph Nutter 
for collaborations~\cite{Nutter:2012an} upon which this brief review is based.

\def\Discussion{
\setlength{\parskip}{0.3cm}\setlength{\parindent}{0.0cm}
     \bigskip\bigskip      {\Large {\bf Discussion}} \bigskip}
\def\speaker#1{{\bf #1:}\ }
\def\endDiscussion{}

\end{document}